\documentclass[12pt]{article}
\usepackage{amssymb}

\usepackage{amsmath}
\usepackage{graphicx}

 \hbox{ \vrule height0pt width5in
\vbox{\hbox{\rm }\break \hbox{UAB-FT-499 \hfill} \hrule height.1cm
width0pt}} \vspace{3mm} \textheight=23.7cm \textwidth=16.5cm
\newcommand{\stackm}{\stackrel{\scriptstyle <}{{ }_{\sim}}}
\newcommand{\stackM}{\stackrel{\scriptstyle >}{{ }_{\sim}}}

\begin{document}

\begin{center}
{\large \textsc{``Steckbrief'' $\lambda$}} \vskip8mm
{\large Joan Sol\`{a}}\footnote{Invited talk at the EuroConference
on Frontiers in Astroparticle Physics and Cosmology, Sant Feliu de
Gu\'{\i}xols, Girona, 30 Sept.-5 Oct. 2000. To be published in
\textsl{Nucl. Phys. B, Proc. Suppl.}, ed. M. Hirsch, G. Raffelt and J.W.F. Valle. } %
\vskip3mm \textsl{Grup de F\'{\i}sica Te\`{o}rica and Institut de
F\'{\i}sica d'Altes Energies (IFAE), }

\textsl{Universitat Aut\`{o}noma de Barcelona, \\ E-08193,
Bellaterra, Barcelona, Catalonia, Spain}
\end{center}

\vspace{1.3cm}

\begin{center}
{ABSTRACT}
\end{center}

\begin{quotation}
The phenomenology of the FLRW models with non-vanishing
cosmological constant, $\lambda$, is briefly surveyed in the light
of the recent astrophysical and cosmological observations. A
subset of these $\lambda\neq 0$  models, which probably includes
the world where we live, is singled out by the combined data from
high redshift Type Ia supernovae, CMBR and the cosmic inventory of
matter in our universe. The kinematical success of a non-vanishing
$\lambda$, however, leaves open many dynamical questions in
quantum field theory. A semiclassical renormalization group
approach to $\lambda$ might perhaps shed some light on them. In
this context, $\lambda$ is naturally non-zero simply because it is
a running parameter.
\end{quotation}

\newpage \vskip12mm
%

\section{Definition of $\lambda$ and a bit of history}

In the following I will give a short review of some classic
matters related to the impact of a non-vanishing cosmological
constant on the kinematical evolution of the universe.  Addressing
the (much harder) issue of the cosmological constant problem in
Quantum Field Theory, is not at all the main purpose of this note.
Still, some of it will be sketched at the very end, hopefully in
pedagogical terms. I will conclude with a remark on the potential
relation of the cosmological constant to the renormalization group
and the role played by the lightest degrees of freedom of our
universe\,\footnote{For a considerably expanded version of this
talk, see Ref.\cite{CChep}. For a classical exposition of the
subject of the cosmological constant problem, see \cite{Weinberg}
and references therein. For an overview of some recent, fairly
advanced, aspects of the problem within the context of string
theory, see Ref.\cite{WittWein} and references therein. For a
detailed discussion of the renormalization group approach to the
cosmological constant problem within the line discussed at the end
of this talk, see Refs.\,\cite{SS1,SS2}.}.

The cosmological constant (CC), $\lambda$,  was first introduced
by Einstein in 1917\,\cite{Sitzung1} two years after he proposed
the gravitational field equations without cosmological
term\,\cite{Sitzung2}. In our conventions\footnote{%
Expert settings: $sign\;\left( g_{\mu \nu }\right) =(+,-,-,-)$; Riemann: $%
R_{\;\mu \nu \eta }^{\lambda }=\partial _{\eta }\Gamma _{\mu \nu
}^{\lambda }-\partial _{\nu }\Gamma _{\mu \eta }^{\lambda }+...$;
Ricci: $R_{\mu \nu }=R_{\;\mu \lambda \nu }^{\lambda }$; Curvature
scalar: $R\equiv g^{\mu \nu }R_{\mu \nu }$.}, the field equations
in the presence of the cosmological constant read
\begin{equation}
G_{\mu \nu }+\lambda g_{\mu \nu }=-8\pi G_{N}T_{\mu \nu }\,,
\label{EE}
\end{equation}
where
\begin{equation}
G_{\mu \nu }\equiv R_{\mu \nu }-\frac{1}{2}g_{\mu \nu }R\,
\label{Gmn}
\end{equation}
is the so-called Einstein's tensor, $T_{\mu \nu }$ is the
energy-momentum tensor and $G_{N}$ is Newton's constant. In
Einstein's words: ``...we may add the fundamental tensor $g_{\mu
\nu }$ multiplied by a universal constant, $-\lambda $ $\left[
{}\right. $in his conventions$\left. {}\right] $, at present
unknown, without destroying the general covariance... This field
equation, with $\lambda $ sufficiently small, is in any case also
compatible with the facts of experience derived from the solar
system... That term is necessary only for the purpose of making
possible a quasi-static distribution of matter, as required by the
fact of the small velocities of the stars''\,\cite{Sitzung1}.

Notice that the field equations (\ref{EE}) in the presence of
$\lambda \neq 0$ can be obtained from the field equations for
$\lambda =0$ with the simple substitution
\begin{equation}
T_{\mu \nu }\rightarrow \tilde{T}_{\mu \nu }=T_{\mu \nu }+\Lambda
\,g_{\mu \nu } \label{Tmunubar}
\end{equation}
We shall see later that the alternative CC parameter $\Lambda
=\lambda /8\pi G_{N}$ plays a role on its own in that it
represents a vacuum energy density. Then Eq.\thinspace (\ref{EE})
can be cast as an effective set of vacuum equations
\begin{equation}
G_{\mu \nu }=-8\pi G_{N}\tilde{T}_{\mu \nu }\,.  \label{EE2}
\end{equation}
Notice furthermore that $\lambda $ and $\Lambda $ are
dimensionful scalar quantities. From the dimensions $\left[
G_{N}\right] =E^{-2}$ and $\left[ T_{\mu \nu }\right] =E^{4}$ in
natural units, we have $\left[ \lambda \right] =E^{2}$ and
$\left[ \Lambda \right] =E^{4}$. Nowadays the most stringent
bounds (or actual estimations) on $\lambda $ come, rather than
from our experience in the solar system, from cosmology (see later
on), and entail
\begin{equation}
\lambda \stackm {\cal O}(10^{-84})\,GeV^{2}\Leftrightarrow \Lambda
\stackm {\cal O}(10^{-47})\,GeV^{4}. \label{CCbound1}
\end{equation}
Since $\lambda $ is not dimensionless, we may assess the dramatic
smallness of this number only by comparing it to another
dimensionful quantity, such as e.g. the bound on the mass
(squared) of the photon, from terrestrial measurements of the
magnetic field:
\begin{equation}
m_{\gamma }^{2}\stackm {\cal O}(10^{-50})\,GeV^{2}\,.
\label{mphoton}
\end{equation}
In spite of our strong and unbreakable faith on the gauge dogma
(which asserts that $m_{\gamma }=0$ exactly!) it turns out that we
happen to know -- by direct experimental knowledge -- that the
(queer and much more unfamiliar) parameter $\lambda $ is many
orders of magnitude smaller than the most believed-to-be-zero
physical parameter in the history of physical science: the photon
mass. \ How could anybody still doubt for a second that $\lambda $
should be zero too? All these compelling pieces of ``evidence''
notwithstanding, that might well not be the case after all!

It is perhaps useful to recall at this point that at the time when
Einstein put forward his field equations with a non-vanishing
cosmological term, astrophysicists did not even know that the
stars that we may watch glowing beautifully in the firmament in a
dark night are members of a very particular galaxy in the
universe, our Milky Way Galaxy, which is just one among a hundred
billions of them scattered amid the voids of the immense cosmos.
Furthermore, from the historical point of view, we should
emphasize that the cosmological constant was introduced by
Einstein as a philosophical fiat, namely one that conformed with
the classical trends of Western's Philosophy. In view of the state
of contemporary knowledge mentioned above, the idea of a static
and ever unchanging cosmos was ingrained very profoundly in the
current culture. Moreover, at the end of the last century it
became notorious Mach's conception (christened ``Mach's
Principle'' by Einstein himself\,\cite{MP}) according to which the
inertia of a body is determined by the bulk matter distribution in
the universe (Mach, 1893). Within this framework one is plausibly
led to the conclusion that the universe had better be finite in
space dimensions otherwise the bodies moving inside could not be
affected by the overall matter content of the world. Another
ingredient supporting the ``necessity'' for a finite universe was
the so-called ``Olbers paradox'', formulated in the middle of the
nineteenth century, namely the fact that if the universe is
infinite and homogeneous (filled with a constant average density
of matter emitting a certain amount of radiation in all
directions) we should observe a permanently bright sky due to the
summing of all the luminous contributions of the uniform matter
distribution in the world. In fact, although the intensity of the
energy irradiated by the shell at distance $r$ dies off as
$1/r^{2}$, the number of stars increases with $r^{3}$, so the net
outcome is an ever increasing amount of radiant energy at all
points of an infinite space. There wouldn't be dark nights to see
the stars!! Of course, according to our modern view, we know that
there is no paradox at all. First, the number of stars within our
physical horizon is, though enormous, finite; with an estimated
total of $\sim 10^{11}$ galaxies, each containing an average of
$\sim 10^{11}$stars, our universe contains, at most, $10^{22}$
shining suns ``filling'' the interstellar medium with an average
density of scarcely a few protons per cubic meter. Second, the
assumption that we receive that\ (approximately) constant amount
of power from every shell is false. While it is reasonable to
think that they do send a similar amount of electromagnetic
energy, it is not true that we accumulate the total. To cure the
disease, there is no need to assume a finite universe, for in an
universe in expansion -- even if infinitely large -- the energy
emitted from farther and farther shells becomes more and more
red-shifted when it reaches the observer, and so the energy
integral can be perfectly finite; in fact, it can be small enough
to allow dark nights with crisp star views!

Hubble's discovery (1929) of the recession of galaxies was the
corner stone setting the experimental standpoint of modern
cosmology. It instantly killed the necessity of a static universe
-- Einstein's original universe-- and, as quoted by
Gamow\,\cite{Gamow}, it made Einstein's to exclaim that the
introduction of the CC in his field equations was ``the biggest
blunder of my life''. As a matter of fact it was no blunder at
all, for it could quite be that the CC is after all a physical
reality -- as I shall discuss in the next sections. Intriguingly
enough, the possibility of a nonvanishing CC is perhaps the most
profound legacy left over by the father of General Relativity.

\section{The $\lambda$-force}

$\lambda $ is indeed a constant independent of the chosen local
inertial frame. The covariant derivative on both sides of
Einstein's Eqs.\,(\ref{EE}) is zero because of local conservation
of energy-momentum ($\nabla ^{\mu }T_{\mu \nu }=0$) and the
automatic (Bianchi) identity $\nabla ^{\mu }G_{\mu \nu }=0$
satisfied by the Einstein tensor. Therefore since $\nabla ^{\mu
}g_{\mu \nu }=0\Longrightarrow \nabla _{\mu }\lambda =0$ and so also $%
\partial _{\mu }\lambda =0\Longrightarrow \lambda $ is a constant scalar
field --i.e. a parameter. To gather a physical intuition on the nature of $%
\lambda $\ in Einstein's equations let us interpret it in terms of
forces. Consider the static gravitational field created by a
source mass $M$ at the origin, with density $\rho _{M}=M\,\,\delta
^{3}(r)$. For weak fields $g_{\mu \nu }\simeq \eta _{\mu \nu }$ is
the usual Lorentz metric. We further assume a non-relativistic
regime where $T_{00}\simeq \rho _{M}$ is just the matter
density of the source. The $(\mu ,\nu )=(0,0)$ component of Eq.\thinspace (%
\ref{EE}) then reads $G_{00}+\lambda =-8\pi G_{N}\,\rho _{M}$ with $%
G_{00}=R_{00}-(1/2)R$. In the low-velocity (non-relativistic) case
we also expect that $T_{ij}\ll T_{00}\,(i,j=1,2,3)$. This is
equivalent to saying that we neglect pressure and stress as
compared to matter density. Therefore, from (\ref{EE}) we are entitled to set $%
G_{ij}+\lambda g_{ij}\simeq 0$. This implies that $R_{ij}\simeq
(1/2R-\lambda )g_{ij}$ and thus the curvature scalar boils down to
$R=g^{\mu \nu }R_{\mu \nu }\simeq R_{00}+3(1/2R-\lambda )$, or $\
R\simeq -2R_{00}+6\lambda $. Substituting this back into the
previous $(0,0)$ field equation we find $R_{00}-\lambda =-4\pi
G_{N}\,\rho _{M}$. Also a very short computation confirms that,
within our approximation, $R_{00}\simeq (-1/2)\nabla ^{2}g_{00}$.
Finally, recalling that Newton's potential $\phi $ is related to
the deviation of the $(0,0)$ component of the metric tensor from
$\ \eta _{00}=1$ (through $\ g_{00}\simeq \ 1+2\phi $) we are led
to the fundamental equation
\begin{equation}
\nabla ^{2}\phi =4\pi G_{N}\left( \rho _{M}-\frac{\lambda }{4\pi G_{N}}%
\right) \,.  \label{Poisson}
\end{equation}
This is nothing but Poisson's equation for the Newton potential
with an
additional term $-\lambda /4\pi G_{N}$ whose sign depends on that of $%
\lambda $. Thus if e.g. $\lambda >0$ the original gravitational
field becomes diminished as though there were an additional
repulsive interaction.
In other words, the sign of the new force $F_{\lambda }$ has the sign of $%
\lambda $. These features are confirmed by explicitly solving Eq.\thinspace (%
\ref{Poisson}):
\begin{equation}
\phi =-\frac{G_{N}\;M}{r}-\frac{1}{6}\lambda r^{2} \label{solPE}
\end{equation}
We are thus led to the expected gravitational potential plus a new
contribution. The additional term is an \ ``harmonic oscillator''
potential -- repulsive for $\lambda >0$!$.$ The corresponding
force field on a test particle of mass $\mu $ ($\mathbf{g=F}/\mu
$) reads
\begin{equation}
{\bf g}=-\frac{G_{N}\;M}{r^{3}}{\bf r}+\frac{1}{3}\lambda {\bf r}
\label{newF}
\end{equation}
and contains an extra linear term with the aforementioned sign
property.

\section{Cosmologies with non-vanishing $\lambda$}

We shall now show the impact of the $\lambda$-force in the
cosmological scenario. We adopt the Cosmological Principle as the
physical paradigm from where to build up our image of the
universe. This principle asserts the isotropy (and \textit{a
fortiori} the homogeneity) of the universe in the large. In turn
this leads to the Friedmann-Lema\^{\i}tre-Robertson-Walker (FLRW)
type of cosmologies. If we concentrate on the matter-dominated
(MD) era (more than $99.99\%$ of the universe lifetime) then the
full spectrum of FLRW models follow from the basic
Friedmann-Lema\^{\i}tre (FL) equation in the presence of a
$\lambda$-term, namely
\begin{equation}
\dot{a}^{2}=\frac{C^{2}}{a}\,-k+\frac{\lambda }{3}\,a^{2}\equiv
F(a)\,\,. \label{FL}
\end{equation}
Here
\begin{equation}
C^{2}\equiv \left( 8\,\pi \,G_{N}\,\,/3\right) \,\rho
_{0}\,a_{0}^{3} \label{C2}
\end{equation}
is a positive constant because its sign is that of $\rho _{0}=\rho
_{M}(t_{0})$, the present energy density of matter.  $F(a)\geq 0$
is, too, by construction a positive-definite function of the scale
parameter $a(t)$. The value of the latter at present is denoted
$a(t_0)\equiv a_{0}$.  Equation (\ref{FL}) follows from Einstein's
field equations (\ref{EE}) if one assumes a matter-energy
distribution respecting the Cosmological Principle, therefore
based on an isotropic energy-momentum tensor $T_{\mu\nu}$.
However, by modeling the universe through an isotropic low-density
medium, e.g. a perfect fluid sphere, pieced together in patches
such that movements (e.g. expansion) are never relativistic in
each piece, a Newtonian approach is possible. Then the energy
conservation law --including the $\lambda$-term from
(\ref{solPE})-- for a patch of mass $\mu$ sitting at $a=a(t)$
reads
\begin{equation}
\frac{1}{2}\,\mu\,\dot{a}^{2}-\rho\left(
\frac{4}{3}\pi\,a^{3}\right)
\frac{G_{N}\,\mu}{a}-\frac{1}{6}\,\lambda
\,\mu\,a^{2}\,=E\,.\label{kinetic}%
\end{equation}
Here $t$ is the cosmological proper time, namely the time measured
by every observer that accompanies the mean motion of the uniform
matter distribution in the universe, modeled as a perfect fluid.
The existence of this cosmic time is a consequence of the
Cosmological Principle. In fact, the latter requires that every
cosmological parameter and field of the ``fluid universe'', be at
most a function of the cosmological proper time, $f=f(t)$, and so
$f$ must be homogeneous and isotropic in space coordinates. In
this respect I should point out that the Cosmological Principle
does not preclude the possibility that the CC could be, as the
scale factor itself, a function of the cosmic time:
$\lambda=\lambda(t)$. From the point of view of the co-moving
observer, this $t$-dependent ``cosmological constant'' is to be
interpreted as an additional gravitational source beyond the
original $T_{\mu\nu}$, in the manner of eq. (\ref{Tmunubar}).
Therefore, the total source is
\begin{equation}
\widetilde{T}_{\mu\nu}=T_{\mu\nu}+\frac{\lambda(t)}{8\pi
G_{N}}\,g_{\mu\nu
}=T_{\mu\nu}+\Lambda(t)\,g_{\mu\nu} \label{lambdat}%
\end{equation}
The difference with respect to eq.(\ref{Tmunubar}) is that in the
present instance it is the total $\widetilde{T}_{\mu\nu}$ that is
conserved, not the original $T_{\mu\nu}$. Of course this is
required by the Bianchi identity satisfied by the Einstein tensor
$G_{\mu\nu}$, as explained at the beginning of Section 2. The case
of a variable $\lambda=\lambda(t)$, however, will not be pursued
here anymore. Let us just mention that it is at the heart of the
so-called ``quintessence'' models\,\cite{Bludman}.

Next consider the equation of continuity of our perfect fluid
medium (in the MD era, where pressure and radiation density are
negligibles in front of matter density $\rho_M$). Since $|{\bf
r}|=a$ and $|\dot{\bf r}|=\dot{a}$ we have $\dot{\bf
r}=\dot{a}\,({\bf r}/a)=H\,{\bf r}$ and hence the equation of
continuity of our non-relativistic fluid model yields
\begin{eqnarray}
&&0=\frac{\partial \rho _{M}}{\partial t}+{\bf\nabla}\cdot (\rho _{M}\,%
{\bf\dot{r}})=\dot{\rho}_{M}+3\,H\,\rho _{M}\nonumber\\
&&\Rightarrow \,\frac{%
d\rho _{M}}{\rho _{M}}+3\,\frac{da}{a}=0\,,  \label{continuity}
\end{eqnarray}
where we notice that, by virtue of the Cosmological Principle,
$\rho _{M}(t)$ can only be a function of $t$ at most. Therefore,
Eq.\thinspace (\ref {continuity}) is trivially integrated to give
$\rho _{M}(t)\,a^{3}(t)=\rho _{0}\,a_{0}^{3}=\mathrm{const.}$.  We
may now substitute this first integral on Eq.(\ref{kinetic}) to
eliminate the density. Moreover, in our Newtonian picture the
Gaussian curvature constant $k$ of the $3$-space sections of the
space-time is recovered through the prescription
\begin{equation}
k=-{\frac{E}{|E|}}\equiv\left\{
\begin{array}
[c]{lll}%
-1 & \mbox{if $E>0$\ (open)} & \\
0 & \mbox{if $E=0$\ (flat)} & \\
+1 & \mbox{if $E<0$\ (closed)} &
\end{array}
\right.\,.\label{kvalues}%
\end{equation}
Of course we are free to rescale the scale factor e.g.
$a\rightarrow (\sqrt{2\left| E\right| /\mu })\,a$. In doing this
Eq.(\ref{kinetic}) finally transforms into the FL equation
(\ref{FL}).

A graphic summary of the FLRW cosmologies is sketched in
Figs.\,1-2. We have divided them into three classes: Class I
($k=-1$), Class II ($k=0$) and Class III ($k=+1$), and each class
subdivides into models depending on the value of the CC. The
analysis of these models is in principle not difficult as the
differential equation (\ref{FL}) can be integrated by quadrature,
\begin{equation}
t_{0}-t=\int_{a(t)}^{a_{0}}\frac{da}{\sqrt{F(a)}}\,.\label{quadrature}%
\end{equation}
From here one obtains $t=t(a)$ and upon inverting one gets
$a=a(t)$. Unfortunately, neither of the last two operations can be
performed analytically in the general case. For $\lambda =0$ (and
any $k$) or for $k=0$ (and any $\lambda $) the integral
(\ref{quadrature}) can be done explicitly, but for $k=\pm 1$ and
$\lambda \neq 0$ it leads to an elliptic function and so numerical
integration is required for an accurate quantitative description.
Nonetheless the qualitative traits of the resulting function
$a=a(t)$, and so the various types of FLRW universes, can be
pinned down analytically in all cases without need of an explicit
numerical analysis.

But before embarking us on further discussions it is convenient to
define the canonical cosmological parameters at the present time.
They are defined to be the present energy density of matter and
cosmological constant in units of the critical density now:
\begin{equation}
\Omega _{M}\equiv \frac{\rho _{0}}{\rho_{0}^{c}}\,,\ \ \ \Omega
_{\Lambda }\equiv \frac{\Lambda }{\rho _{0}^{c}}=\frac{\lambda
}{3\,H_{0}^{2}}\,,  \label{cpparameters}
\end{equation}
\begin{equation}
\rho ^{c}_{0} \equiv \frac{3\,H_{0}^{2}}{8\,\pi \,G_{N}} =\left(
3.0\,\sqrt{h_{0}}\times 10^{-3}\,eV\right) ^{4}\,. \label{rhocrit}
\end{equation}
Here the dimensionless number $h_{0}\sim 0.65\pm 0.1$\,\cite{Silk}
sets the typical range for today's value of Hubble's ``constant''
\begin{equation}
H_{0}\equiv\left({\dot{a}\over a}\right)_0=100\;{Km/sec\over
Mpc}\;h_{0}\,. \label{HubblesC}
\end{equation}
From these parameters the FL Eq.(\ref{FL}) can be trivially cast
in the form of an exact sum rule for the present time:
\begin{equation}
\Omega _{M}+\Omega _{\Lambda }+\Omega _{K}=1\,,
\label{cosmictriangle}
\end{equation}
where $\Omega_{K}=-k/a_{0}^{2}\,H_{0}^{2}$ is the cosmological
curvature parameter, which is seen to be dependent of the other
two previously defined. In the absence of $\lambda $ the resulting
cosmological models are extremely simple: Cf. Models I2, II2 and
IIIb in Figs.\,1-2. However, these are just very particular cases
of the full collection of $14$ FLRW models with $\lambda \neq 0$
displayed in these figures. We remark that FLRW models with
vanishing CC have the property that the universe is spatially open
($k=-1$), closed ($k=+1$) or flat --i.e. Euclidean ($k=0$)-- if
and only if it expands forever, ultimately re-collapses (into a
``Big Crunch'' point) or expands just up to the border between
expansion and re-collapse, respectively. Nevertheless such a
one-to-one correspondence between the ultimate destiny of the
universe and the topological structure of its associated $3$-space
no longer holds when $\lambda \neq 0$. For example, a spatially
closed universe (hence a compact one) with non-vanishing CC could
well be one that expands forever (Fig.\thinspace 2), and a
spatially open or flat universe with non-vanishing CC could
ultimately recollapse (Fig.\thinspace 1).
\begin{figure*}[tb]
\begin{tabular}{clr}
\mbox{\hspace{1.75cm}} & \resizebox{!}{10cm}{\includegraphics{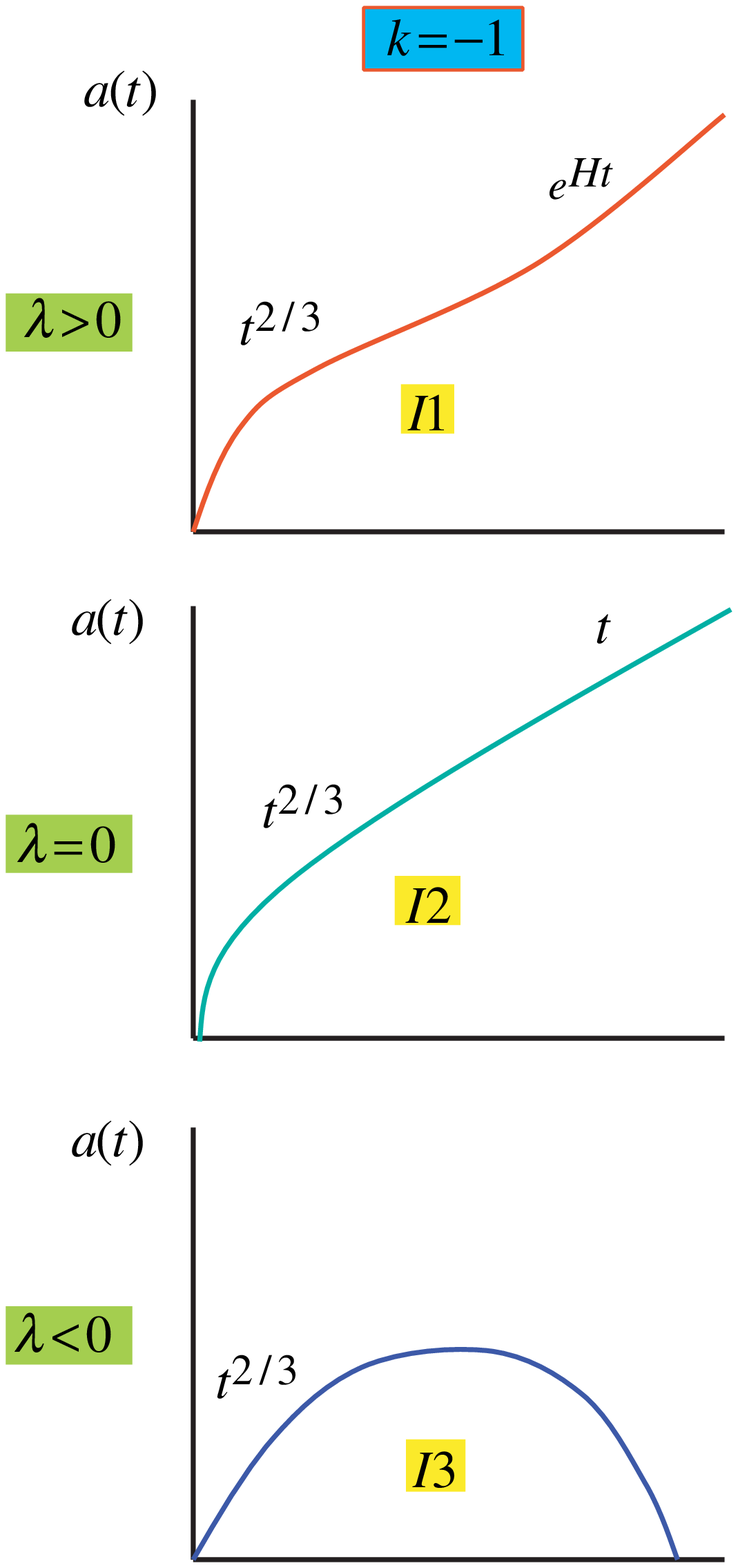}} & %
\resizebox{!}{10cm}{\includegraphics{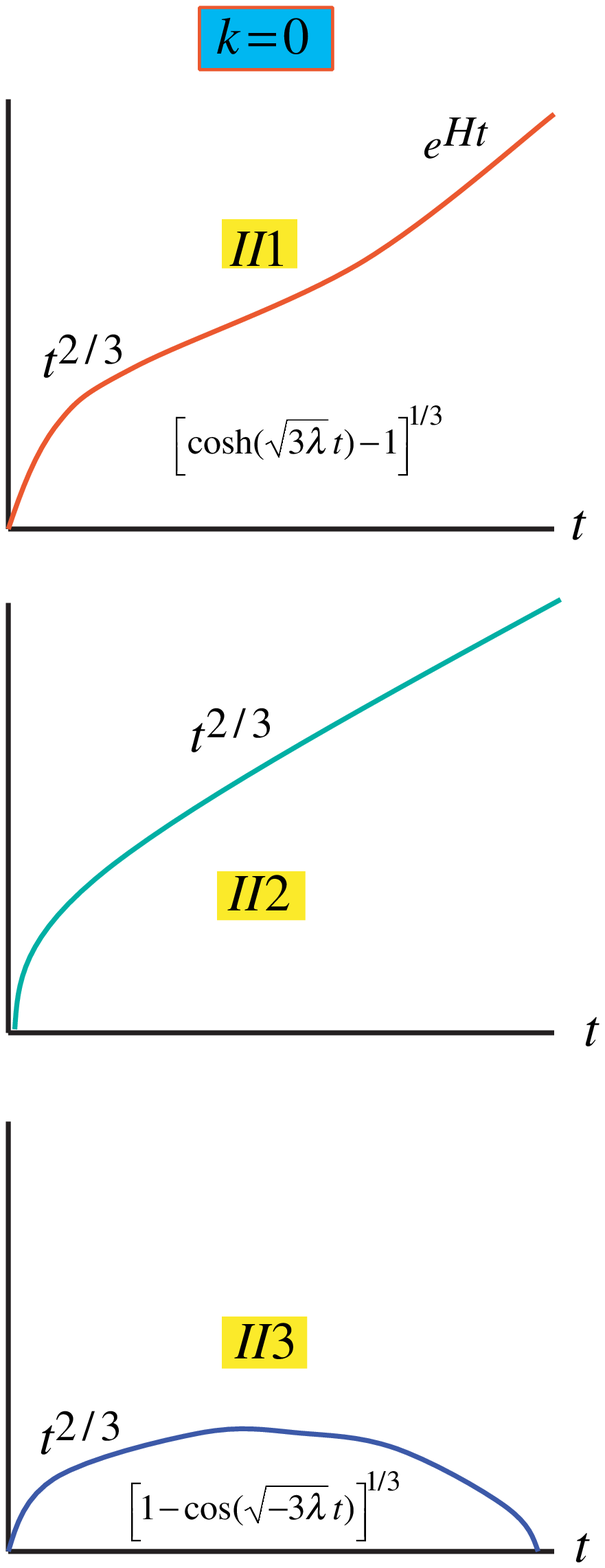}}
\end{tabular}
\caption{The Class I and Class II FLRW models. Model II1 is both
theoretically and experimentally favored. Model II2 is the
Einstein-de Sitter model.} \label{fig:f2ab}
\end{figure*}
\begin{figure*}[tb]
\begin{tabular}{clr}
\mbox{\hspace{1cm}} & \resizebox{!}{10cm}{\includegraphics{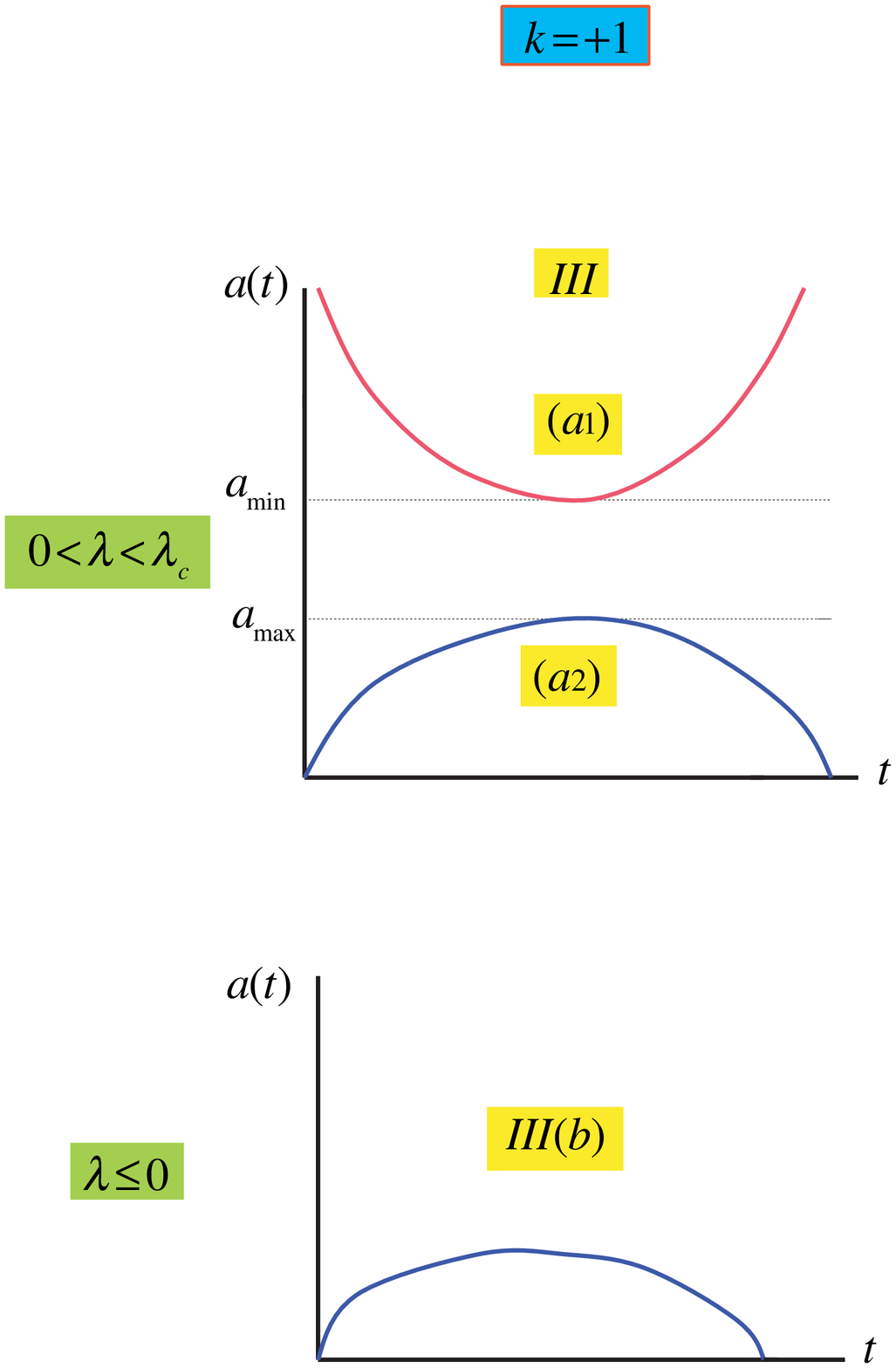}} & %
\resizebox{!}{10cm}{\includegraphics{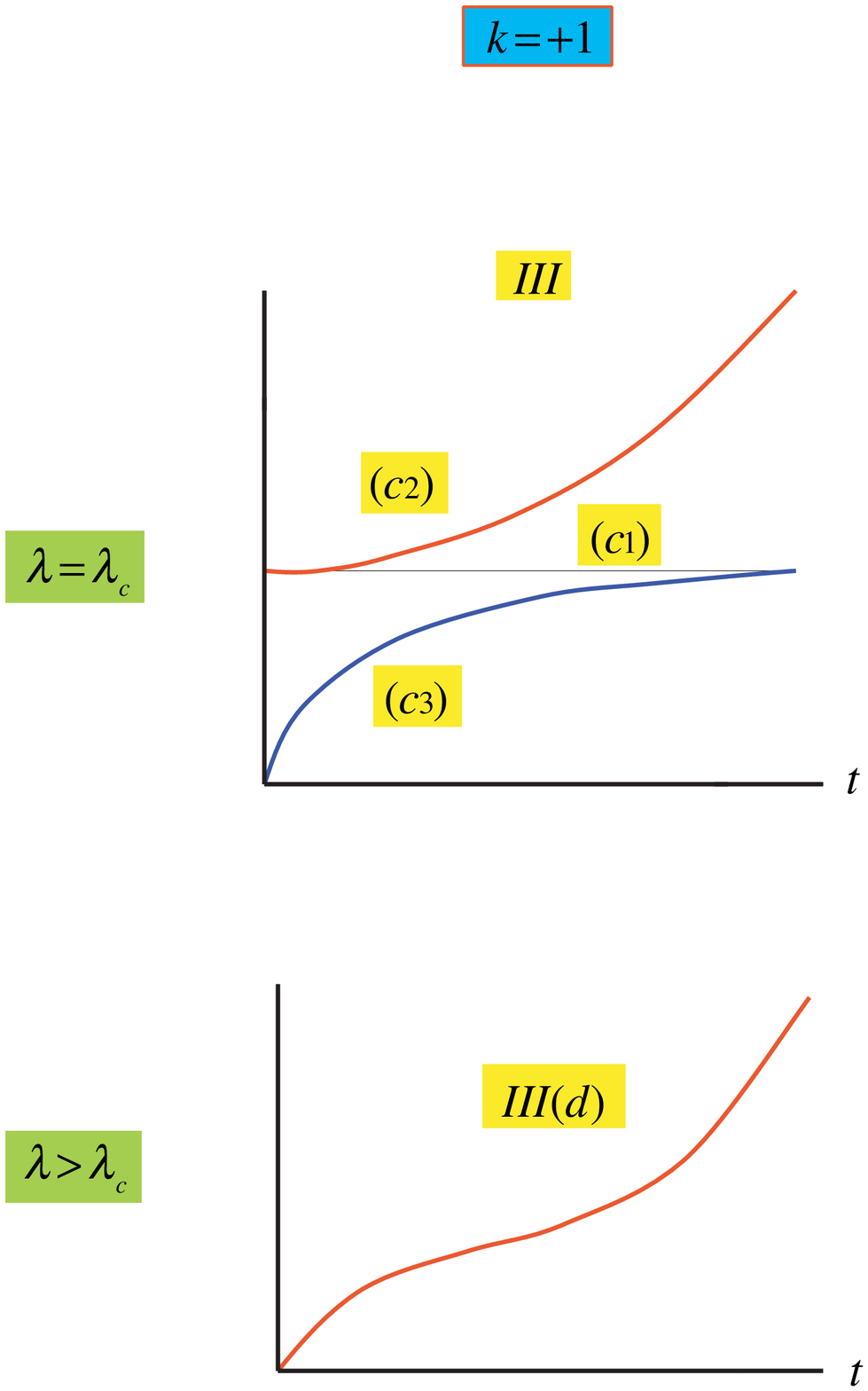}}
\end{tabular}
\caption{The rich Class III of FLRW models. In particular, Model
III(c1) is Einstein's static model, Model III(c2) is the
Eddington-Lema\^{\i}tre model and Model III(d) is similar to Model
II1, except that $k=+1$. It could also be a favored case, see
Fig.\,3a.} \label{fig:f3ab}
\end{figure*}
In the following I wish to comment a bit more on a few of the FLRW
models in Figs.1-2. Model II2 is a $\lambda=0$ universe of
particular historical interest, viz. the Einstein-de Sitter (EdS)
model from 1932, just devised by these authors after Einstein
abjured in 1931 the creature he had engendered fourteen years
before: the CC itself! The EdS model is the simplest FLRW
cosmological model accounting for the observed expansion. However,
it is a ``critical model'', namely the density of matter is
exactly equal to the critical density, $\rho_0=\rho_0^c$
(equivalently, $\Omega_M=1$), and for this reason
eq.(\ref{cosmictriangle}) trivially implies that it is a spatially
flat universe with vanishing CC.  The EdS model solves in a very
simple analytical form upon integrating Eq.(\ref{quadrature}) with
$k=0$ and $\lambda=0$, with the result:
$a^{3}(t)=(3\,C\,t/2)^{2}$. Notice that the behavior $a(t)\propto
t^{2/3}$, valid for all $t$ only in this model, is nevertheless
characteristic of all the FLRW models with initial singularity
(``Big Bang'') when we approach $t\rightarrow 0$ within the MD
epoch, see Figs.\,1-2. Paradoxically, the EdS model is nowadays
excluded (see Fig.\,3a) as it gives a really poor man fit to the
combined data from high redshift Type Ia
supernovae\,\cite{Perlmutter}, the temperature anisotropies in the
cosmic microwave background radiation
(CMBR)\,\cite{Silk,JaffeBahcall,Bernardis} and the dynamical
observation of clustered matter\,\cite{JaffeBahcall}. At the end
of the day the (formerly abhorred) CC is back again, and with
renewed momentum!  Models with positive CC are singled out by
observation at the $99\%$ C.L.!

Particularly favored is Model II1 (Cf. Fig.\,1). In this case one
can also derive from (\ref{quadrature}) the exact analytical
evolution of the scale factor for all $t$. It involves a
hyperbolic cosine function whose asymptotic (exponential) behavior
matches that of the closed model I1. The result is
\begin{equation}
a^{3}(t)=\frac{3\,C^{2}}{2\,\lambda }\,\ \left[ \cosh \left( \sqrt{%
3\,\lambda }\,t\right) -1\right]\,.  \label{ut}
\end{equation}
For convenience I have normalized this solution such that
$a(0)=0$. We verify that $a(t)\sim\exp {(H_{\lambda }\,t)}$ for
$t\rightarrow \infty $, where $H_{\lambda }\equiv
\sqrt{\lambda/3}$  is Hubble's constant in the de Sitter space.

Class III is the one with the richest spectrum of models, see
Fig.\, 2. There is a critical value $\lambda
_{c}\equiv 4/(9\,C^{4})>0$ 
obtained from requiring that the discriminant of the cubic
equation $F(a)=0$ is zero in order to guarantee a double
(positive) real root $a=a_{c}\equiv 3C^{2}/2=1/\sqrt{\lambda
_{c}}$. The subclasses are easily identified. If $0<\lambda
<\lambda _{c}$
(see Fig.2, up-left), then there is an excluded segment $a_{\mathrm{max}}<a(t)<a_{%
\mathrm{min}}$ in which $F(a)<0$. There are two allowed models of
this sort depending on whether $0\leq a(t)\leq a_{\mathrm{max}}$
--Model III(a2)-- or $a(t)>a_{\mathrm{min}}$ --Model III(a1).
Notice that the former model is oscillatory whereas the latter
(``bouncing universe'') has the curious property that it has no
$t=0$ singularity. In fact, Model III(a1) contracts exponentially
from infinity and then expands back towards
the same place at the same pace. As for the two cases $\lambda <0$ and $%
\lambda =0$ in Fig.\thinspace 2 (down-left), they are both dubbed
Model III(b) and are similar to the oscillatory models discussed
before. Worth noticing are the three subclasses IIIc ($\lambda
=\lambda _{c}$) in Fig.\thinspace 2 (up-right). The remarkable
thing about Model III(c1) is that it is just the original static
Einstein model at fixed $a=a_{c}$. Though static, it corresponds
to an unstable fixed point of the FL differential equation
(\ref{FL}). This was already noticed by Eddington on simple
physical grounds: if for some reason this universe would expand
slightly, this would diminish the gravitational attraction but at
the same time would enhance the repulsive $\lambda $-force
($\lambda _{c}>0$!) \ because the latter is larger the larger is
the separation between particles. As a result the original ``seed
expansion'', no matter how small it is, would destabilize the
universe into a ``runaway expansion''. Similarly, an initial
``seed contraction'' would cause the universe to shrink
indefinitely. Mathematically, by perturbing the solution one
obtains, in one case, a non-singular model -- Model III(c2)--
which starts at $a=a_{c}$ and then evolves exponentially up to
infinity (the so-called Eddington-Lema\^{\i}tre model), and in the
other case a model -- Model III(c3)-- which starts at the
singularity $a=0$ and then creeps up asymptotically towards
$a=a_{c}$. Finally, for $\lambda
>\lambda _{c}$ one obtains Model III(d) in Fig.\thinspace 2 (down-right)
which is similar to models I1 and II1. Interestingly enough,
if Model III(d) is such that $\lambda $ is only slightly larger than $%
\lambda _{c}$, then there appears an approximately flat region in
the middle of the curve and we obtain a quasi-Einstenian model in
which $a(t)$ loiters a long while around $a=a_{c}$ before the
eventual (exponential) de Sitter's phase takes over. It follows
that, for an appropriate choice of $\lambda >0$, such a
``Lema\^{\i}tre's hesitation universe'' can be made arbitrarily
old!

\section{The universe where we live}

After this short review of the FLRW cosmologies with non-vanishing
CC we are now in position to discriminate between the most favored
models according to the latest experimental observations. As
already mentioned, the flat and critical ($\Omega_M=1$) EdS model
(Model II2), which was a preferred cosmological scenario for about
40 years (viz. through a period mediating from 1932 until the
$70´s$), is no longer favored. As a matter of fact it is deadly
ruled out by the combined data (Cf. Fig.\,3a). Although the EdS
model is a prototype dark matter model, it turns out to predict
too much dark stuff!!  For, as can be seen in Fig.\,3a, the most
recent galaxy clustering observations seem to point towards a
low-density universe ($\Omega_M<1$). This fact together with the
supernova data insisting on a positive CC of the order (actually
larger than that) of the matter density gives a final verdict
excluding the flat EdS universe. There is, however, another flat
model, although certainly a non-critical one, in our list of Sec.
3, which is nowadays a most cherished candidate for a viable model
of our universe: Model II1 in Fig.\,2. This FLRW model is a flat
universe with positive CC ($k=0\,,\lambda
>0$). It is strongly highlighted both on
theoretical grounds (by the inflationary paradigm) and
experimentally -- because it is compatible with the supernovae
data, the astrophysical inventory of clustered matter, the revised
age determinations of the globular clusters (the oldest objects
known in our galaxy) and also with the precise measurements of the
temperature anisotropies in the CMBR\,\cite{Silk}. Thus at present
we have a consistent solution to the various ``age problems''
plaguing this field in the past. To fix this conundrum it helps to
have a non-vanishing and positive CC, but also the fact that the
revised ages of the globular clusters are smaller than previously
thought \,\cite{Chaboyer}.
\begin{figure*}[tb]
\begin{tabular}{ccc}
\mbox{\hspace{1cm}} & \mbox{\hspace{1cm}} (a) & \mbox{\hspace{0.7cm}} (b) \\
\mbox{\hspace{1cm}} &
\resizebox{!}{7.2cm}{\includegraphics{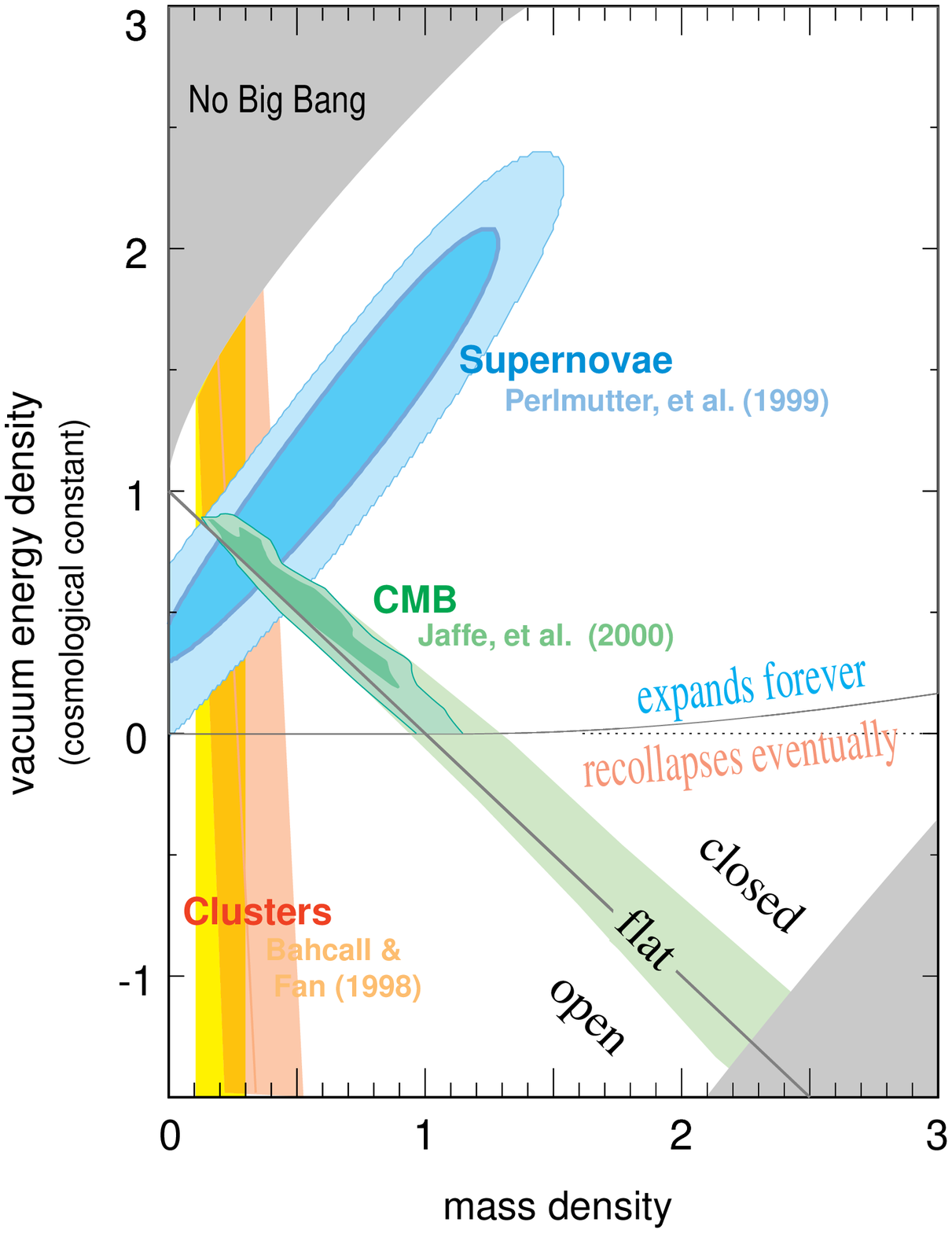}} & \resizebox{!}{6.7cm}{%
\includegraphics{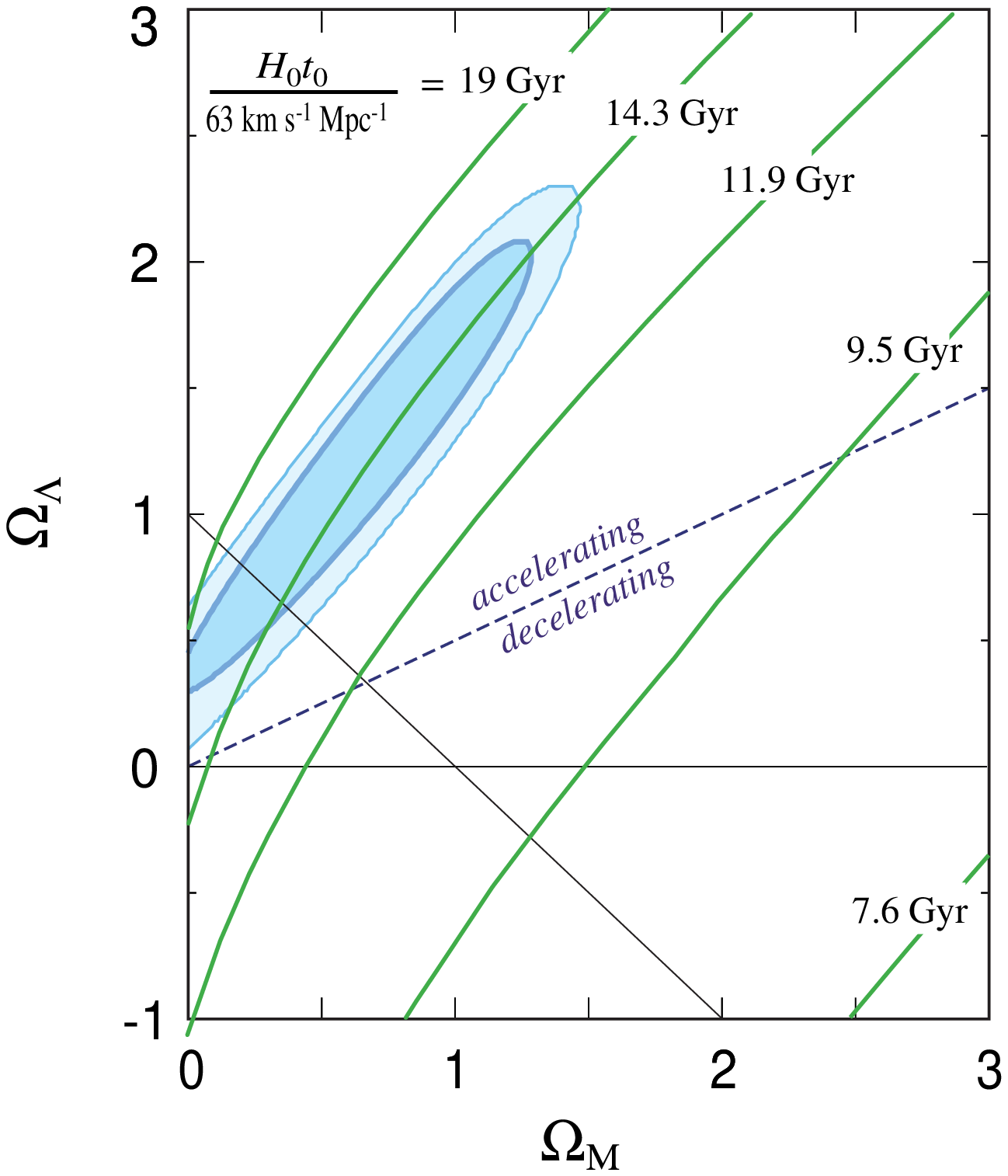}}
\end{tabular}
\caption{\textbf{(a)} Combined data from high redshift Type Ia
supernovae, CMBR and clustered matter observations plotted in the
plane $(\Omega _{M},\Omega _{\Lambda })$ defined by the mass
density and the vacuum density (CC) in units of the critical
density. It is seen that the $\protect\lambda >0$ FLRW
cosmological models with $0<\Omega _{M}<\Omega _{\Lambda }$
expanding forever are clearly favored; \textbf{(b)} Type Ia
supernovae data as in (a) showing that the preferred model is an
accelerating universe about $14-15\,Gyrs$ old. See Refs.
\protect\cite{Perlmutter,Ariel1,JaffeBahcall,Bernardis}.}
\label{fig:f4ab}
\end{figure*}
The best candidate model universe, Model II1, is singled out as a
crossover area in Fig.\,3a around the flat space point
$\Omega_{M}\simeq0.28,\,\Omega_{\Lambda}\simeq+\,0.72$. In the
vicinity of this point the supernovae data, the CMBR data and the
dynamical counting of clustered matter in the universe are best
met than in any other region of the $(\Omega _{M},\Omega _{\Lambda
})$ plane. Notwithstanding, we should emphasize that the data just
off the $k=0\,,\lambda
>0$ line ($\Omega_M+\Omega_{\Lambda}=1$) in Fig.\,3a is tilted
into the domain of closed, low-mass, universes with positive CC.
Therefore, from the strict point of view of the experimental
observation, a closed universe of the Type III(d) in Fig.\,2
cannot be excluded in spite of all the theoretical prejudices that
we might have in mind!

Worth noticing is that none of the non-singular (i.e. no Big Bang)
FLRW cosmological models is singled out by supernovae data and
CMBR (Cf. Fig.\, 3). For example, the shaded set of points on the
left upper corner of Fig. 3a correspond to ``bouncing universes''
-- see Model\ III(a1) in Fig.\,2 -- namely those that shrink down
to a minimum from infinity and then recede to that point in the
future. These universes are seen to be the oldest ones (Cf.
Fig.\,3b) but cannot be accepted, and not only because they are
incompatible with supernovae data. But also because it can be
proven that at the point of highest shrinking the density
parameter is bounded from above by a too small quantity
\begin{equation}
\Omega_{M}\leq \frac{2}{z_{\rm max}^2(z_{\rm max}+3)}\stackm 0.01
\end{equation}
whose numerical value reflects the fact that we have already
observed objects (e.g. quasars) with redshifts $z\stackM 5$ -- or
even higher according to very recent, preliminary, data on remote
proto-galaxies. On the other hand, the formerly (very famous)
Lema\^{\i}tre ``loitering universes'' lying on the border line
around the set of ``bouncing universes'' (upper left corner of
Fig.3a) are seen to be also excluded. As for the shaded set of
points on the down-right corner of Fig. 3a, it is also ruled out
(see Fig. 3b) on the grounds that these universes are too young
($t_{0}<9\,Gyr$) so that the oldest heavy elements would not have
had time to form. In short, in the light of the present
cosmological observations the existence of a tiny $\lambda
>0$ does help in an essential way to reconcile the age of the
universe with the age of the oldest (known) objects living inside
it and in general to obtain an overall picture which is in harmony
with the experimental reckoning of cosmic matter and relic
radiation.

\section{Cosmological Constant and Particle Physics}

\subsection{The CC problem in the SM}

In spite of the goodness of a non-vanishing CC from the point of
view of cosmological kinematics, the existence of a tiny positive
cosmological constant poses serious dynamical questions that go to
the heart of Theoretical Elementary Particle Physics inasmuch as
it is based on Quantum Field Theory (QFT). For instance, there are
enormous contributions to the CC in the Standard Model (SM) coming
from the spontaneous breaking of the electroweak symmetry, i.e.
from the vacuum expectation value of the Higgs potential. If we
call $\Lambda _{ind}$ the (overall) induced vacuum energy density
from QFT, then the total energy-momentum tensor gets an additional
vacuum contribution: $T_{\mu \nu }\rightarrow T_{\mu \nu
}+<0|T_{\mu \nu }|0>$ in which, by Lorentz covariance, $<0|T_{\mu
\nu }|0>=g_{\mu \nu }\,\Lambda _{ind}$ (Cf. Eq.(\ref{Tmunubar}) )
Therefore, in the semiclassical approach, the total effective
cosmological constant entering Einstein's equations (\ref{EE})
reads
\begin{equation}
\lambda _{eff}=\lambda +8\,\pi \,G_{N}\,\Lambda _{ind}\equiv
\lambda +\lambda _{ind}\,.  \label{efflambda}
\end{equation}
Of course it is this effective quantity (the sum of the ``vacuum
CC'' and the ``induced CC'') what the supernovae and CMBR
measurements must have pinned down. Then the ``cosmological
constant problem''\,\cite{Weinberg} appears in the context of the
SM when one considers the contribution from the QCD and
electroweak vacuum energies to $\lambda _{ind}$ in
Eq.(\ref{efflambda}). Let us focus on the electroweak part only.
The Higgs potential of the SM reads
\begin{equation}
V_{cl}=-\frac{1}{2}m^{2}\phi ^{2}+\frac{f}{8}\phi ^{4}. \label{2a}
\end{equation}
Let $v$ be the vacuum expectation value (VEV) of $\phi$, namely
the value where $V_{cl}$ becomes minimum. Then, shifting the
original field $\phi \rightarrow H^{0}+v$ such that the physical
scalar field $H^{0}$ has zero VEV, one obtains the physical mass
of the Higgs boson: $M_{H}=\sqrt{2}\;m$. At the minimum of the
potential (\ref{2a}):
\begin{equation}
\phi =\sqrt{\frac{2m^{2}}{f}}=v\,\,\,\,\,\,{\rm and}\,\,\,\,\,\,\,f=\frac{%
M_{H}^{2}}{v^{2}}\,.  \label{5N}
\end{equation}
From (\ref{5N}) one obtains the following value for the potential,
at the tree-level, that goes over to the induced CC:
\begin{equation}
\Lambda _{ind}=<V_{cl}>=-\frac{m^{4}}{2f}\,.  \label{nnn6}
\end{equation}
If we apply the current numerical bound $\,M_{H}\stackM 115\,GeV$
from LEP II, then the corresponding value $\left| \Lambda
_{ind}\right| \simeq 1.0\times 10^{8}\,GeV^{4}$ is $55$ orders of
magnitude greater than the observed CC from the supernovae.
(Recall that $\Lambda _{ph}\sim 10^{-47}GeV^{4}$ --Cf.
Eq.(\ref{CCbound1})-- as it follows from using the favored value
$\Omega_{\Lambda}\sim 0.7$ mentioned in Sec. 4, and $\rho_{0}^c$
given in Eq.(\ref{rhocrit})). Clearly, unless we fine tune the
original $\lambda$ term (or ``vacuum CC term'') on the RHS of
Eq.(\ref{efflambda}) with a precision of $55$ decimal places we
are in trouble. But of course, even if doing this fantastic fine
tuning, which is technically possible in principle, we are still
in trouble because it has to be repeated order by order in
perturbation theory, and this is certainly untenable.

\subsection{A remark on a renormalization group approach}

Many theoretical ideas have been proposed to solve this
ever-growing conundrum \,\cite{Weinberg}. However, for lack of
space, I will only mention the possibility, recently put forward
in Refs.\cite{SS1,SS2}, that the CC has to be treated as a running
parameter in a semiclassical formulation of the gravitational
field equations. This way does not provide the fundamental
solution of the CC problem either. Nevertheless it helps in better
understanding the problem and (maybe even more important) in
drawing some physical consequences out of it. The basic idea is
that in QFT the vacuum action is subject to renormalization and to
the renormalization group running. At any given energy scale $\mu$
the CC will have a different value $\Lambda(\mu)$ driven by a
renormalization group equation (RGE) and boundary conditions.
Consequently, the ``cosmological constant'' is not a constant,
still less should be zero. In this approach one can derive the
contributions from the light particles to the running of the
cosmological and gravitational constants in the SM, starting from
the cosmic scale up to the Fermi scale. In particular, near the
present cosmic scale $\mu\sim\mu_c$ (see below) one can show that
the full CC (induced plus vacuum parts) obeys a RGE driven by the
lightest degrees of freedom (d.o.f.) available in the universe.
The final form of the RGE for the physical CC depends not only on
the RGE for the vacuum term, but also on the RGE for the induced
term, i.e. the RGE for the VEV of the effective potential, eq.
(\ref{nnn6}). However, the latter is already determined by the RGE
of the SM couplings and parameters. In the general case, the
relevant equation is given in \,\cite{SS2}, but at the present
epoch of our universe it just boils down to \,\cite{SS1}
\begin{equation}
(4\pi )^{2}\frac{d\Lambda _{ph}}{dt}\,=\frac{1}{2}\,m_{S}^{4}\,-\,4\,%
\sum_{i}\,m_{\nu _{i}}^{4}\,. \label{MasterRGE}
\end{equation}
Here $S$ is a very light scalar field (which we may call
``Cosmon'')\footnote{This name was first minted in Ref.\cite{PSW}
and then used also in \cite{CW} and \cite{JSP}. While it should be
emphasized that the present approach is completely different, the
name is kept because the final aim is of course the same.}, whose
mass $m_S$ is a few times the average mass of the lightest
neutrinos $m_j$ \,\cite{Valle}. Typically these may include the
electron neutrino and a sterile neutrino. Let us just consider the
electron type; then $m_S\stackM 4\,m_{\nu _{e}}$\,\cite{SS1}.
Thanks to the scalar nature of the Cosmon, the running can be such
that if one starts with zero CC at the very far infrared (IR)
epoch of the universe, a positive CC can be generated at the
present time and with the right order of magnitude according to
the supernovae experiments. In addition, this new point of view
helps to get a grasp to the so-called ``cosmic coincidence
cosmological constant problem'', namely the problem of why the
measured CC just happens to be of the order of the present day
matter density. This is tantamount to asking why the CC starts to
dominate the energy density of the universe at the epoch of
structure formation. Of course one can invoke ``anthropic
considerations''\,\cite{WittWein}, but from our point of
view\,\cite{SS2} the value of $\Lambda$ at present is obtained
from RG arguments alone once the initial conditions are fixed at
some renormalization point. The latter can typically be the very
far IR scale, $\mu_{IR}$, where the value of the CC can in
principle be whatever. The scale $\mu_{IR}$ is the ``ultimate
energy scale'' down the present day cosmic scale $\mu_{c}$ where
no active d.o.f. are available -- from the RG point of view. It is
reasonable to associate the cosmic scale at present with a
quantity of order of (Cf. eq.(\ref{rhocrit}))
\begin{equation}\label{muc}
\mu_c\sim\left(\rho_0^c\right)^{1/4}=3.0\,\sqrt{h_{0}}\times
10^{-3}\,eV\,.
\end{equation}
This scale is near the value of the lightest neutrino masses
invoked to solve the various neutrino puzzles\,\cite{Valle}.
Therefore, one can arrange for the Cosmon and the lightest
neutrino to be the only RG-active d.o.f. at present, as explained
above. Then the RGE for the physical CC in the segment from the
far IR up to the scale of the next-to-lightest-neutrino, say
$\nu_{\mu}$, is given as follows:
\begin{equation}
(4\pi )^{2}\frac{d\Lambda _{ph}}{dt}\,=\left\{
\begin{array}{ccc}
\frac{1}{2}\,m_{S}^{4}\,-\,4\,m_{\nu_e}^{4} & \mbox{ $(m_{\nu_{\mu}}>\mu>m_S>m_{\nu_e})$\ } &  \\
? & \mbox{ $(\mu_{IR}\leq\mu<m_{\nu_e})$\ } &
\end{array}
\right. \,.  \label{newfor}
\end{equation}
Here we have normalized $t$ such that $t=\ln (\mu /\mu _{IR})$. In
general we can expect $\mu _{IR}\ll \mu _{c}$. The value $\Lambda
_{ph}(IR)$ of the CC in this ``ultimate'' energy scale (where no
active d.o.f. remain) can be zero or not, but in any case we do
not know the running near it because we ignore if there are extra
(ultralight) d.o.f. in its immediate vicinity. If, however, one
assumes (perhaps by invoking some string
symmetry\,\cite{WittWein}) that $\Lambda(\mu_{IR})=0$, and that
there are no other d.o.f. than those already considered, then the
value $\Lambda_{ph}(\mu)$ at any scale $\mu>\mu_{IR}$ becomes
determined, and in particular also the value $\Lambda_{ph}(\mu_c)$
at the present cosmic scale $\mu_c$. In this case we have the
following interesting situation. On the one hand
$\Lambda_{ph}(\mu_c)$ is just obtained from the RGE
(\ref{newfor}), implying that the present day physical value of
the CC is, roughly, $\Lambda_{ph}\equiv\Lambda_{ph}(\mu_c)\sim
m_S^4\stackM m_{\nu_e}^4$. And, on the other hand, we have
$\rho_0^c\sim\left((2-3)\times 10^{-3}\,eV\right)^4$, and this
number happens to be of order $m_{\nu_e}^4$ -- and of course of
order $m_{S}^4$. Hence one obtains the desired relation
$\Lambda_{ph}\sim\rho_0$,
which ``explains'' the supernovae data and the ``cosmic
coincidence''. It should be pointed out that, within our
framework, this relation is equally valid now as it was in the
epoch of structure formation. This is because, as mentioned above,
the next-to-lightest d.o.f. ready to contribute to the RHS of the
RGE (\ref{MasterRGE}) is the muon (and tau) neutrino, which in the
canonical solutions to the neutrino puzzles\,\cite{Valle} are
nearly degenerate and of order $1\,eV$. Therefore, since they are
three orders of magnitude heavier than the lightest neutrino, they
should have already decoupled from the RG evolution of the
physical CC at the time of structure formation, and so
eq.(\ref{newfor}) still applies at that time. This justifies our
contention. Finally, it is worth mentioning that when
extrapolating the running of the CC at higher energies in this
framework (e.g. energies of order of the electron mass, in which
the electron becomes an RG-active d.o.f.) one can show that the
scaling dependences of the cosmological (and gravitational)
constants do not spoil primordial nucleosynthesis, a quite
rewarding result \,\cite{SS2}.

{\bf Acknowledgments}: I am grateful to the organizers for the
kind invitation and the nice atmosphere met in Sant Feliu de
Gu\'{\i}xols. I thank S. Bludman, \ M.C. Gonz\'{a}lez-Garc\'{\i}a,
A. Goobar, I.L. Shapiro and J.W.F. Valle for useful conversations.
I thank also A. Goobar for providing me Fig. 3a. This work has
been supported in part by CICYT under project No. AEN99-0766.

\end{document}